\documentclass[pra,showpacs,twocolumn]{revtex4}
\usepackage{dcolumn}
\usepackage{bm}
\usepackage{amssymb}
\usepackage{amsmath}
\usepackage{amsfonts}

\usepackage[]{graphicx}
\begin{document}
\bibliographystyle{prsty}
\title{A dual-Lagrangian description adapted to quantum optics in dispersive and dissipative dielectric media}
\author{ Aur\'elien Drezet $^{1}$}
\address{(1) Univ. Grenoble Alpes, CNRS, Institut N\'{e}el, F-38000 Grenoble, France
}
\begin{abstract}
We develop a dual description of quantum optics adapted to dielectric systems without magnetic property. Our formalism, which is shown to be equivalent to the standard one within some dipolar approximations  discussed in the article, is applied to the description of polaritons in dielectric media. We show that the dual formalism leads to the Huttner-Barnett equations [B. Huttner, S. M. Barnett, Phys. Rev. A \textbf{46}, 4306 (1992)] for QED in dielectric systems. More generally, we discuss the role of electromagnetic duality in the quantization procedure for optical systems and derive the structure of the dynamical laws in the various representations.     
\end{abstract}

 \maketitle
\section{Introduction}
\indent  The description of quantized electromagnetic fields propagating in a dielectric media has been the subject of intensive research since the early times of quantum optics. It was shown already  in 1948 \cite{Jauch1948} that an Hamiltonian formalism adapted to non dispersive and dissipative model can be easily implemented  by analogy with standard QED quantization procedures used in vacuum. With the development of quantum optics and integrated photonics several authors have extended the previous canonical quantization method to complex inhomogeneous systems with a spatially dependent  dielectric permittivity $\varepsilon(\mathbf{x})$ (see for example Refs.~\cite{Knoll1987,Glauber1991,Wubs2003}). However, in order to consider the physical properties of realistic optical media the central issue is the inclusion of dispersion and dissipation in the quantum description~\cite{Drumond1990}. Since these properties are connected by the well-known Kramers-Kronig relations~\cite{Jackson1999,Lifshitzbook} related to causality it is clear that a correct description  should not analyze dispersion and dissipation separately. Nevertheless, when dissipation is weak it is actually possible to give an Hamiltonian basis to the quantum formalism~\cite{Constantinou1993,Milonni1995,Al1996,Garrison2004} using relations obtained long ago by L. Brillouin~\cite{Lifshitzbook,Brillouin} for the energy density in dispersive media at frequency $\omega$, at which absorption is negligible. These approaches involve `quasi-modal' expansions and were recently applied to the field of quantum plasmonics~\cite{Tame2013} for the coupling of emitters to nano antennas~\cite{Waks2010,Yang2015,Archambault2010} when dissipation is weak. Clearly these assumptions are however not generally valid, e.g., in plasmonics~\cite{Raether,Barnes,Agio}, where optical resonances whith low quality factors are coupled to fluorescent emitters~\cite{Chang2006,Akimov,Fedutik,Wei,Kolesov,Schie,Schell,CucheNL2010,MolletPRB2012,Berthel2015}.\\ 
\indent However, in the 1990's it already became clear that fundamental progresses could only occur by including more degrees of freedom in the quantum description, i.e., by modeling dissipation with thermal baths coupled to the photonic variables. The approach was used in quantum optics for the description of lossy beam splitters~\cite{Jeffers1993,Barnett1998}, for instance for modeling Casimir forces~\cite{Genet2003}, and more recently was applied in quantum plasmonics by modeling decay channels as an information loss in the metal/dielectric environment~\cite{Tame2008,Ballester2009,Ballester2010}.
Moreover, the most important progress came in 1992 when it was suggested by Huttner and Barnett to model any dispersive and absorbing linear bulk media satisfying Kramers-Kronig relations by using a modified Hopfield-Fano~\cite{Fano1956,Hopfield1958} model involving a continuous distribution of harmonic oscillator fields interacting with light~\cite{Huttner1991,Huttner1992a,Huttner1992b,Huttner1992c,Matloob1995,Matloob1996,Barnett1995}. This powerful approach was originally limited to infinite and homogeneous bulk media but continuous efforts have been done later on to extend its range of validity to the more interesting inhomogeneous medium case, which is needed for nanophotonics applications in general and for quantum plasmonics in particular~\cite{Wubs2001,Suttorp2004a,Suttorp2004b,Suttorp2007,Philbin2010}. Actually these formal developments of the original Huttner-Barnet model~\cite{Huttner1992a} are strongly motivated by the parallel development of the phenomenological Green function approach of Gruner and Welsch~\cite{Yeung1996,Gruner1995,Gruner1996} based on the Langevin equation method~\cite{Cohen} and the dyadic Green function formalism~\cite{Girard1996,Novotny}. In this strategy a noise current is added phenomenologically to Maxwell's equation in order to preserve unitarity of the full evolution and in particular the constancy of all conjugate canonical variables commutator with time~\cite{Scheel1998}. The Green function strategy has been intensively used in the literature in the recent years~\cite{Dung1998,Dung2000,Scheel2001,Matloob1999,Matloob2004,Fermani2006,Raabe2007,Amooshahi2008,Scheelreview2008} and applied to several problems including dielectric or magnetic materials, and coupling of light with atoms in the regime of weak or strong coupling in presence of plasmonic nanoparticles~\cite{Dzotjan2010,Cano2011,Hummer2013,Chen2013,Delga2014,Hakami2014,Choquette2012,Grimsmo2013,Rousseaux2016}. Despite its success the Langevin noise method applied to macroscopic electrodynamics (unlike for atomic physics in vacuum~\cite{Cohen}) lacks a neat and clear quantum foundation that, like the Huttner-Barnett model~\cite{Huttner1992a}, could be justified using an Hamiltonian description. In Refs.~\cite{Wubs2001,Suttorp2004a,Suttorp2004b,Bhat2006,Judge2013,Philbin2010} general proofs were given for the formal equivalence between the Huttner Barnett model and the Langevin noise approach.\\
\indent The present work contributes to this discussion by providing a different Lagrangian and Hamiltonian foundation to the approach considered in, e.g., Ref.~\cite{Philbin2010}. More precisely, our aim is to bypass the usual canonical procedure based on the  `minimal coupling' Lagrangian in the Coulomb gauge~\cite{Cohen2}. Generally speaking this is usually done in the literature by introducing the Power-Zienau unitary transformation~\cite{Zienau,Cohen2} leading to the multipolar representation of the electromagnetic field~\cite{Craig}. Here, we proceed differently by introducing an equivalent description of the electromagnetic system using a dual electric vector potential $\textbf{F}(\mathbf{x},t)$ different from the usual magnetic vector potential $\textbf{A}(\mathbf{x},t)$. With this formalism there is no unphysical separation between transverse quantized and longitudinal unquantized fields (indeed, this separation is no relativistically causal~\cite{Cohen2}). The method of quantization we propose  here is actually a direct way to introduce the multipolar Hamiltonian~\cite{Craig,Zienau}, where the fundamental electromagnetic quantities are now the displacement $\textbf{D}$ and magnetic $\textbf{B}$ fields. In turn this approach, adapted to neutral dielectric systems without magnetic properties, will shed some new light on the Huttner-Barnett model (which was based on the minimal coupling scheme) and therefore on the derivation of the Langevin Noise method~\cite{Yeung1996,Gruner1995,Gruner1996}. We conclude the article by a discussion concerning electromagnetic duality between electric and magnetic physical quantities and  compare  different but equivalent representations of the dynamical laws.       
\section{The general Huttner-Barnett effective model for a linear dielectric medium }
\subsection{A Lagrangian formulation}
We here follow the standard approach in quantum electrodynamics and start with writing a Lagrange-Hamilton principle adapted to our system. In other words, we choose the Lagrangian $L(t)$ and the Lagrangian density $\mathcal{L}(\mathbf{x},t)$ such that the variational problem 
$\delta\int_{t_1}^{t_2}dt L(t)= \delta\int_{t_1}^{t_2}dt\int d^3\mathbf{x}\mathcal{L}(\mathbf{x},t)=0$ could be solved and lead to the Maxwell equations in a dissipative and dispersive inhomogeneous dielectric medium. However, in this work we will not consider the usual canonical `minimal coupling' Lagrangian in the Coulomb gauge \cite{Cohen} which was used historically by Hutner and Barnett~\cite{Huttner1992a} but instead work with the dual choice (in the Lorentz-Heaviside system of units):
\begin{eqnarray}
\mathcal{L}=\frac{\mathbf{B}^2-\mathbf{D}^2}{2}+ \mathbf{F}\cdot\boldsymbol{\nabla}\times\mathbf{P} -\frac{\mathbf{P}^2}{2}+\mathcal{L}_M.\label{1}
\end{eqnarray}
Here, the Lagrangian density depends on the electric vector potential $\mathbf{F}(\mathbf{x},t)\in \mathbb{R}^3$ and its derivatives  defined as 
\begin{eqnarray}
\mathbf{B}(\mathbf{x},t)=\frac{1}{c}\partial_t\mathbf{F}(\mathbf{x},t), &\mathbf{D}(\mathbf{x},t)=\boldsymbol{\nabla}\times\mathbf{F}(\mathbf{x},t).\label{3}
\end{eqnarray}
$\mathbf{B}(\mathbf{x},t)$ and $\mathbf{D}(\mathbf{x},t)$ being the magnetic and displacement fields respectively. Additionally here we will work exclusively in the `Coulomb' gauge $\boldsymbol{\nabla}\cdot\mathbf{F}(\mathbf{x},t)=0$. As detailed in the Appendix A and discussed in the section IV the  Lagrangian is motivated by fundamental duality relations existing between electric and magnetic variables. This in turn explains why the magnetic field term  $ \textbf{B}^2/2$ appears with a positive sign in Eq.~1 while $\textbf{D}^2/2$ appears with a negative sign contrarily to the usual choice made in the minimal coupling representation. We point out that this duality symmetry was discussed long ago in a more classical context involving for example chirality or bi-anisotropic materials~\cite{Stratton,Serdyukov,Corbaton2013}, here we applied this concept to the Lagrange-Hamilton language adapted to second quantization.  Moreover, in this formalism, the material part $\mathcal{L}_M$ reads
\begin{eqnarray}
\mathcal{L}_M=\int_{0}^{+\infty}d\omega\frac{(\partial_t\mathbf{X}_\omega)^2-\omega^2\mathbf{X}_\omega^2}{2}.\label{2}
\end{eqnarray} 
where $\mathbf{X}_\omega(\mathbf{x},t)$ defines the material oscillator vectorial fields which are coupled to the electromagnetic variables. These  fields are labeled by the pulsation $\omega$ varying continuously from zero to $+\infty$. We emphasize that in the total Lagrangian density we have also a term $-\frac{\mathbf{P}^2}{2}$ which plays the role of an internal interaction and which is not included in $\mathcal{L}_M$ to respect the conventions used in Ref.~\cite{Huttner1992a}.  This choice for $\mathcal{L}$ leads  to the same dynamical equations as those deduced from the usual formalism based on the minimal coupling Lagrangian~\cite{Cohen}. However, the dual formalism appears more convenient for the present study of neutral dielectrics system since it does not actually involve a non physical separation of the electromagnetic field between a transverse (quantized) part associated with photons and a longitudinal (un-quantized) part  associated with non-causal Coulomb Fields (see later on this paper).  The present description can actually be seen as an alternative  way to access to the multipolar representation which is connected to the minimal coupling one  through  a unitary transformation as shown by Power and Zienau~\cite{Zienau,Craig,Cohen}. Here the use of the $\textbf{F}$ potential vector allows us to deduce similar results by introducing different canonical variables.     
From the previous definitions follow directly (i.e., using the Euler-Lagrange formalism) the two Maxwell equations: 
\begin{eqnarray}
\boldsymbol{\nabla}\times\mathbf{B}(\mathbf{x},t)=\frac{1}{c}\partial_t\mathbf{D}(\mathbf{x},t), &\boldsymbol{\nabla}\cdot\mathbf{D}(\mathbf{x},t)=0.\label{4}
\end{eqnarray}
The matter field $\mathbf{X}_\omega(\mathbf{x},t) \in \mathbb{R}^3$ is related to the polarization density by the formula
\begin{eqnarray}
\mathbf{P}(\mathbf{x},t)=\int_{0}^{+\infty}d\omega\sqrt{\frac{2\sigma_\omega(\mathbf{x})}{\pi}}\mathbf{X}_\omega(\mathbf{x},t)\label{5}
\end{eqnarray}
where the coupling function $\sigma_\omega(\mathbf{x})\geq 0$ will represent the conductivity of the medium at the harmonic pulsation $\omega$. The fact that the conductivity is sufficient for describing the field will be shown to be consistent with Kramers-Kronig relations~\cite{Huttner1992a}. We emphasize that the  part $\mathcal{L}_M$ of the Lagrangian is not exactly the same as the one used in Hutner-Barnett's theory in which three contributions associated with the electromagnetic field, the dielectric and the oscillator bath were included. Here, we use the simpler model proposed by Philbin~\cite{Philbin2010}  where the electromagnetic field characterized in our approach by $\textbf{F}$ is directly coupled to the oscillator bath variables $\mathbf{X}_\omega$ without including the additional and non necessary mechanical oscillator used in refs.~\cite{Huttner1991,Huttner1992a,Huttner1992b,Huttner1992c,Matloob1995,Matloob1996,Barnett1995}.\\
\indent From Eqs.~\ref{1},\ref{2} we deduce straightforwardly the Euler-Lagrange equations for the electromagnetic fields, i.e., the two missing Maxwell equations:
\begin{eqnarray}
\boldsymbol{\nabla}\times\mathbf{E}(\mathbf{x},t)=-\frac{1}{c}\partial_t\mathbf{B}(\mathbf{x},t), &\boldsymbol{\nabla}\cdot\mathbf{B}(\mathbf{x},t)=0\label{6}
\end{eqnarray}
with the electric field \begin{eqnarray}\mathbf{E}(\mathbf{x},t)=\mathbf{D}(\mathbf{x},t)-\mathbf{P}(\mathbf{x},t).\label{7}\end{eqnarray} 
For the matter oscillator field $\mathbf{X}_\omega(\mathbf{x},t)$ we similarly obtain the second-order Euler-Lagrange equation 
\begin{eqnarray}
\partial_t^2\mathbf{X}_\omega(\mathbf{x},t)+\omega^2\mathbf{X}_\omega(\mathbf{x},t)=\sqrt{\frac{2\sigma_\omega(\mathbf{x})}{\pi}}\mathbf{E}(\mathbf{x},t),\label{8}
\end{eqnarray}
 which contains a linear coupling to the local electric field (there is no magnetic coupling such as a Lorentz force). The term $-\frac{\mathbf{P}^2}{2}$ in the Lagrangian density Eq.~\ref{1} is necessary for deriving this dynamical equation and obtaining a coupling proportional to  $\textbf{E}(\mathbf{x},t)$ and not to $\textbf{D}(\mathbf{x},t)$ (inversely, in absence of coupling with $\textbf{D}(\mathbf{x},t)$ the electric field in Eq.~\ref{8} is $-\textbf{P}(\mathbf{x},t)$ which can be seen as an internal force acting on the oscillators $\mathbf{X}_\omega(\mathbf{x},t)$).\\
\indent Furthermore, in order to use the Hamilton formalism needed for quantization we define the canonical momenta associated with the different fields:
\begin{eqnarray}
\mathbf{\mathbf{\Pi}}_\mathbf{X_\omega}=\frac{\delta \mathcal{L}}{\delta(\partial_t\mathbf{X}_\omega)}=\frac{\partial \mathcal{L_\omega}}{\partial(\partial_t\mathbf{X}_\omega)}=\partial_t\mathbf{X}_\omega\nonumber\\
\mathbf{\mathbf{\Pi}}_\mathbf{F}=\frac{\partial \mathcal{L}}{\partial(\partial_t\mathbf{F})}=\frac{\mathbf{B}}{c}\label{9}
\end{eqnarray}
where $\frac{\delta[...]}{\delta(\partial_t\mathbf{X}_\omega) }$ is a functional derivative and where $\mathcal{L}=\int_0^{+\infty}d\omega\mathcal{L}_\omega$.  This allows us to introduce the full Hamiltonian defined as $H=\int d^3\mathbf{x}[\mathbf{\mathbf{\Pi}}_\mathbf{F}\partial_t\mathbf{F}+\int_0^{+\infty}d\omega\mathbf{\mathbf{\Pi}}_\mathbf{X_\omega}\partial_t\mathbf{X}_\omega)]-L$. We get: 
\begin{eqnarray}
H(t)=\int d^3\mathbf{x}[\frac{\mathbf{B}^2+\mathbf{D}^2}{2}- \mathbf{F}\cdot\boldsymbol{\nabla}\times\mathbf{P} +\frac{\mathbf{P}^2}{2}]+H_M\nonumber\\
=\int d^3\mathbf{x}[\frac{\mathbf{B}^2+\mathbf{D}^2}{2}-\mathbf{D}\cdot\mathbf{P} +\frac{\mathbf{P}^2}{2}]+H_M\nonumber\\
=\int d^3\mathbf{x}\frac{\mathbf{B}^2+\mathbf{E}^2}{2}+H_M\nonumber\\ \label{10}
\end{eqnarray}
with the helpful condition: $\int d^3\mathbf{x}\mathbf{D}\cdot\mathbf{P}=\int d^3\mathbf{x}\mathbf{F}\cdot\boldsymbol{\nabla}\times\mathbf{P}$, and where the material contribution $H_M(t)$ reads:
\begin{eqnarray}
H_M(t)=\int d^3\mathbf{x}\int_{0}^{+\infty}d\omega\frac{(\partial_t\mathbf{X}_\omega)^2+\omega^2\mathbf{X}_\omega^2}{2}.\label{11}
\end{eqnarray}
For the purpose of canonical quantization the full Hamitonian can be written as a functional of the conjugate canonical variables and we deduce:
 \begin{eqnarray}
H(t)=\int d^3\mathbf{x}[\frac{c^2\mathbf{\mathbf{\Pi}}_\mathbf{F}^2+\boldsymbol{\nabla}\times\mathbf{F}^2}{2}- \mathbf{F}\cdot\boldsymbol{\nabla}\times\mathbf{P} +\frac{\mathbf{P}^2}{2}]\nonumber\\
+\int d^3\mathbf{x}\int_{0}^{+\infty}d\omega\frac{(\mathbf{\mathbf{\Pi}}_\mathbf{X_\omega})^2+\omega^2\mathbf{X}_\omega^2}{2}
\nonumber\\ \label{10b}
\end{eqnarray}   
At this stage it is useful to deduce Eq.~\ref{10} using a different approach. Indeed, the Hamiltonian used here is an integral form of the local energy-tensor conservation law associated with Noether's theorem and it can be preferable, for the sake of  generality, to use such a local approach instead of a global one.  Starting from Maxwell's equation obtained with Eqs.~\ref{1}, \ref{2} we  get the Poynting  conservation theorem
\begin{eqnarray}
-\partial_t(\frac{\mathbf{B}^2+\mathbf{E}^2}{2})=\boldsymbol{\nabla}\cdot(c\mathbf{E}\times\mathbf{B})+\mathbf{J}\cdot\mathbf{E}\label{12},
\end{eqnarray}  which involves the local electric current $\mathbf{J}=\partial_t\mathbf{P}$ associated with the polarization density $\mathbf{P}$. Replacing $\mathbf{P}$ by its value in Eq.~\ref{5} and after direct integration leads to $\mathbf{J}\cdot\mathbf{E}=\partial_t(\int_{0}^{+\infty}d\omega\frac{(\partial_t\mathbf{X}_\omega)^2+\omega^2\mathbf{X}_\omega^2}{2})$ and therefore to the local conservation law \begin{eqnarray}
-\partial_t u=\boldsymbol{\nabla}\cdot(c\mathbf{E}\times\mathbf{B})\label{13},
\end{eqnarray}  where $u$, the local energy density, is given by  $u=\frac{\mathbf{B}^2+\mathbf{E}^2}{2}+\int_{0}^{+\infty}d\omega\frac{(\partial_t\mathbf{X}_\omega)^2+\omega^2\mathbf{X}_\omega^2}{2}$. Integration over an infinite volume leads to the Hamiltonian $H(t)=\int d^3\mathbf{x}u(\mathbf{x},t)$  which becomes equivalent to the total conserved energy if the Poynting vector flow $\oint_{\Sigma_\infty} c\mathbf{E}\times\mathbf{B}\cdot d\Sigma$ over the infinite closed surface $\Sigma_\infty$ surrounding our system vanishes sufficiently well.  
\subsection{From the Lagrangian to the polarisability of the linear dielectric medium}
We should now summarize briefly the consequence of the Lagrangian choice Eq.~\ref{1} and show that it allows us to justify the usual Maxwell equations in a causal dielectric medium satisfying Kramers-Kronig relations.   
To do that we have to solve our dynamical coupled equations for matter and electromagnetic fields. In this section we start with the material field which is the easiest part. In order to solve the evolution Eq.~\ref{8} we introduce here other field variables:
\begin{eqnarray}\mathbf{Z}^{(\pm)}_\omega(\mathbf{x},t)=\partial_t\mathbf{X}_\omega(\mathbf{x},t)
\pm i\omega\mathbf{X}_\omega(\mathbf{x},t)\nonumber\\=(\mathbf{Z}^{(\mp)}_\omega(\mathbf{x},t))^\ast\label{14}
\end{eqnarray}
which obey to the following first order equations:
\begin{eqnarray}
\partial_t\mathbf{Z}^{(\pm)}_\omega(\mathbf{x},t)=\pm i\omega\mathbf{Z}^{(\pm)}_\omega(\mathbf{x},t)+\sqrt{\frac{2\sigma_\omega(\mathbf{x})}{\pi}}\mathbf{E}(\mathbf{x},t).\label{15}
\end{eqnarray}
These equations are easily solved using the method of the variation of constants and lead to 
\begin{eqnarray}
\mathbf{Z}^{(\pm)}_\omega(\mathbf{x},t)=\mathbf{Z}^{(\pm)}_\omega(\mathbf{x},t_0)e^{\pm i\omega(t-t_0)}\nonumber\\+\sqrt{\frac{2\sigma_\omega(\mathbf{x})}{\pi}}\int_0^{t-t_0}d\tau e^{\pm i\omega\tau}\mathbf{E}(\mathbf{x},t-\tau)\nonumber\\
=\mathbf{Z}^{(\pm,0)}_\omega(\mathbf{x},t)\nonumber\\+\sqrt{\frac{2\sigma_\omega(\mathbf{x})}{\pi}}\int_0^{t-t_0}d\tau e^{\pm i\omega\tau}\mathbf{E}(\mathbf{x},t-\tau)\label{16}
\end{eqnarray}
with $t_0$ an initial time which eventually could be sent infinitely in the remote past, i.e., if $t_0\rightarrow-\infty$. We also introduced the `free' solution $\mathbf{Z}^{(\pm,0)}_\omega(\mathbf{x},t)=\mathbf{Z}^{(\pm)}_\omega(\mathbf{x},t_0)e^{\pm i\omega(t-t_0)}$ corresponding to the harmonic oscillation of the medium in the absence of coupling.\\
\indent We can of course go back to the initial field variables by using the transformation $\mathbf{X}_\omega(\mathbf{x},t)=\frac{\mathbf{Z}^{(+)}_\omega(\mathbf{x},t)}{2i\omega}+cc.$ and 
$\partial_t\mathbf{X}_\omega(\mathbf{x},t)=\frac{\mathbf{Z}^{(+)}_\omega(\mathbf{x},t)}{2}+cc.$ This leads to 
\begin{eqnarray}
\mathbf{X}_\omega(\mathbf{x},t)
=\mathbf{X}^{(0)}_\omega(\mathbf{x},t)\nonumber\\+\sqrt{\frac{2\sigma_\omega(\mathbf{x})}{\pi}}\int_0^{t-t_0}d\tau \frac{\sin{\omega\tau}}{\omega}\mathbf{E}(\mathbf{x},t-\tau)\label{17}
\end{eqnarray} with the `free' solution $\mathbf{X}^{(0)}_\omega(\mathbf{x},t)=\frac{\mathbf{Z}^{(+,0)}_\omega(\mathbf{x},t)}{2i\omega}+cc.$ (i.e., $\mathbf{X}_\omega^{(0)}(\mathbf{x},t)
=\cos{(\omega (t-t_0))}\mathbf{X}_\omega(\mathbf{x},t_0)+\frac{\sin{(\omega (t-t_0))}}{\omega}\partial_t\mathbf{X}_\omega(\mathbf{x},t_0)$). The polarization density now reads 
\begin{eqnarray}
\mathbf{P}(\mathbf{x},t)=\mathbf{P}^{(0)}(\mathbf{x},t)+\int_{0}^{t-t_0}\chi(\mathbf{x},\tau)d\tau\mathbf{E}(\mathbf{x},t-\tau)\label{18}
\end{eqnarray}
with the free dipole density distribution:
\begin{eqnarray}
\mathbf{P}^{(0)}(\mathbf{x},t)=\int_{0}^{+\infty}d\omega\sqrt{\frac{2\sigma_\omega(\mathbf{x})}{\pi}}\mathbf{X}^{(0)}_\omega(\mathbf{x},t)\label{19}
\end{eqnarray}
and the linear susceptibility:
\begin{eqnarray}
\chi(\mathbf{x},\tau)=\int_{0}^{+\infty}d\omega\frac{2\sigma_\omega(\mathbf{x})}{\pi}\frac{\sin{\omega\tau}}{\omega}\Theta(\tau)\label{20}
\end{eqnarray} For convenience we introduced the Heaviside unit step function $\Theta(\tau)$, defined as $\Theta(\tau)=1$ unless $\tau<0$ whereas $\Theta(\tau)=0$. Now, we get for the displacement  field
 \begin{eqnarray}
\mathbf{D}(\mathbf{x},t)=\mathbf{E}(\mathbf{x},t)+\mathbf{P}(\mathbf{x},t)=\mathbf{P}^{(0)}(\mathbf{x},t)\nonumber\\+\mathbf{E}(\mathbf{x},t)+\int_{0}^{t-t_0}d\tau\chi(\mathbf{x},\tau)\mathbf{E}(\mathbf{x},t-\tau).\label{21}
\end{eqnarray}
If as usually we consider the limit  $t_0\rightarrow-\infty$ we can write \begin{eqnarray}
\mathbf{D}(\mathbf{x},t)=\mathbf{P}^{(0)}(\mathbf{x},t)+\int_{-\infty}^{+\infty}d\tau\frac{\varepsilon(\mathbf{x},\tau)}{2\pi}\mathbf{E}(\mathbf{x},t-\tau)\label{22}
\end{eqnarray}
where the dielectric permittivity of the polarizable medium is defined as
\begin{eqnarray}
\frac{\varepsilon(\mathbf{x},\tau)}{2\pi}=\delta(\tau)+\chi(\mathbf{x},\tau).\label{23}
\end{eqnarray}
In the following we will keep this useful definition even if we do not work in the limit $t_0\rightarrow-\infty$. \\
\indent  At this stage it is useful to introduce the Fourier transform of the fields defined as $\widetilde{A}(\omega)=\int_{-\infty}^{+\infty}\frac{dt}{2\pi}e^{i\omega t} A(t)$ and $ A(t)=\int_{-\infty}^{+\infty}d\omega e^{-i\omega t}\widetilde{A}(\omega)$. If we temporally suppose that it makes mathematically sense to define a Fourier transform for  $\mathbf{D}(\mathbf{x},t)$, $\mathbf{E}(\mathbf{x},t)$and $\mathbf{P}(\mathbf{x},t)$ then in the  $t_0\rightarrow-\infty$ limit of Eq.~\ref{22} we get 
\begin{eqnarray}
\widetilde{\mathbf{D}}(\mathbf{x},\omega)=\widetilde{\mathbf{P}}^{(0)}(\mathbf{x},\omega)+\widetilde{\varepsilon}(\mathbf{x},\omega)\widetilde{\mathbf{E}}(\mathbf{x},\omega)\nonumber\\
\widetilde{\varepsilon}(\mathbf{x},\omega)=1+2\pi\widetilde{\chi}(\mathbf{x},\omega)\label{24}
\end{eqnarray} 
Actually it is not at all obvious that such a Fourier transform can be defined unambiguously.  Fluctuating and stationary quantum fields are not in general converging sufficiently fast in the future or past directions so that the Fourier transform is not in general a well defined mathematical object for real frequency $\omega$. The problem can be more rigorously handled using Laplace's transform~\cite{Wubs2001,Suttorp2004a}. We will not consider this problem in this article keeping the complete analysis for a future work. Here it is sufficient to observe that in optics the Fourier transform  
\begin{eqnarray}
\widetilde{\varepsilon}(\mathbf{x},\omega)=1+\int_0^{+\infty}d\tau\chi(\mathbf{x},\tau)e^{i\omega \tau} \label{25}
\end{eqnarray} is an analytical function in the upper part of the complex plane $\omega=\omega'+i\omega''$, i.e., $\omega''>0$, provided $\chi(\mathbf{x},\tau)$ is finite for any time $\tau\geq 0$.
From this naturally follows the symmetry $\widetilde{\varepsilon}(\mathbf{x},-\omega)^\ast=\widetilde{\varepsilon}(\mathbf{x},\omega^\ast)$. From these analytical properties it is possible to derive the general Kramers-Kronig relations existing between  the real part $\textrm{Re}[\widetilde{\varepsilon}(\mathbf{x},\omega)]\equiv\widetilde{\varepsilon}'(\mathbf{x},\omega)$ and the imaginary part  $\textrm{Imag}[\widetilde{\varepsilon}(\mathbf{x},\omega)]\equiv\widetilde{\varepsilon}''(\mathbf{x},\omega)$ of the permittivity. Remarkably Eq.~\ref{20} fully satisfies these conditions. Indeed, we can write the Fourier transform of Eq.~\ref{20} in the upper complex plane  \begin{eqnarray}
\widetilde{\chi}(\mathbf{x},\omega'+i\omega'')=\int_{0}^{+\infty}\frac{du}{\pi}\frac{\sigma_{u}(\mathbf{x})/\pi}{u^2-(\omega'+i\omega'')^2}\nonumber\\
=\int_{-\infty}^{+\infty}\frac{du}{2\pi}\frac{\sigma_{u}(\mathbf{x})}{\pi u}\frac{1}{u-\omega'-i\omega''}\label{26}
\end{eqnarray}
where we used the relation:
\begin{eqnarray}
\frac{1}{u^2-(\omega'+i\omega'')^2}=\frac{1}{2u}[\frac{1}{u-\omega'-i\omega''}+\frac{1}{u+\omega'+i\omega''}]\nonumber\\ \label{27}
\end{eqnarray} as well as the symmetry (definition): $\sigma_{u}(\mathbf{x})=\sigma_{-u}(\mathbf{x})$.
We thus get in the limit $\omega''\rightarrow 0^+$ 
\begin{eqnarray}
\widetilde{\varepsilon}(\mathbf{x},\omega'+i0^+)=1+\int_{-\infty}^{+\infty}\frac{du}{\pi}\frac{\sigma_{u}(\mathbf{x})}{u}\frac{1}{u-\omega'-i0^+}\nonumber\\
=1+P\left(\int_{-\infty}^{+\infty}\frac{du}{\pi}\frac{\sigma_{u}(\mathbf{x})}{u}\frac{1}{u-\omega'}\right)+i\frac{\sigma_{\omega'}}{\omega'}.\nonumber\\ \label{28}
\end{eqnarray} where $P[...]$ denotes Cauchy's principal value. 
This is Kramers-Kronig relation if we use the identity $\widetilde{\varepsilon}''(\mathbf{x},\omega')=\frac{\sigma_{\omega'}(\mathbf{x})}{\omega'}$ along the real axis. We have thus 
$\widetilde{\varepsilon}''(\mathbf{x},\omega)=-\widetilde{\varepsilon}''(\mathbf{x},-\omega)$ and $\widetilde{\varepsilon}'(\mathbf{x},\omega)=\widetilde{\varepsilon}'(\mathbf{x},-\omega)$ in agreement with the symmetry requirement (from Eq.~\ref{26} we deduce $\widetilde{\chi}(\mathbf{x},-\omega)^\ast=\widetilde{\chi}(\mathbf{x},\omega^\ast)$). Eq.~\ref{20}  characterizing the generalization of Hutner-Barnett model~\cite{Huttner1992a} is thus a complete representation of a causal linear dielectric medium including both dispersion and dissipation. 
\section{The Formal quantization procedure}
\subsection{Quantizing the matter field equations}
\indent In order to obtain a quantized theory of the matter field it is useful to introduce the auxiliary fields $\mathbf{f}^\ast_\omega(\mathbf{x},t)$, $\mathbf{f}_\omega(\mathbf{x},t)$  which  correspond to rising-lowering operators:
\begin{eqnarray}
\sqrt{\frac{\hbar}{2\omega}}\mathbf{f}_\omega(\mathbf{x},t)=\frac{\mathbf{Z}^{(-)}_\omega(\mathbf{x},t)}{-2i\omega},
\sqrt{\frac{\hbar}{2\omega}}\mathbf{f}^\ast_\omega(\mathbf{x},t)=\frac{\mathbf{Z}^{(+)}_\omega(\mathbf{x},t)}{2i\omega}\nonumber\\ \label{28}
\end{eqnarray}
with  $\mathbf{X}_\omega(\mathbf{x},t)=\sqrt{\frac{\hbar}{2\omega}}(\mathbf{f}_\omega(\mathbf{x},t)+\mathbf{f}^\ast_\omega(\mathbf{x},t))$. From this we can obtain the 
following representation for the Hamiltonian $H_M$:
\begin{eqnarray}
H_M=\int d^3\mathbf{x}\int_{0}^{+\infty}d\omega\hbar\omega\mathbf{f}^\ast_\omega(\mathbf{x},t)\mathbf{f}_\omega(\mathbf{x},t) \label{30}
\end{eqnarray}
Canonical quantization starts with the replacement of the vector fields $\mathbf{f}_\omega$ by operators acting on the Hilbert space associated with the quantum system under study. In particular, we have the replacement $\mathbf{f}^\ast_\omega(\mathbf{x},t)\rightarrow\mathbf{f}^\dagger_\omega(\mathbf{x},t)$. 
The equal-time commutators between the conjugate canonical variables read:
\begin{eqnarray}
[\mathbf{X}_\omega(\mathbf{x},t),\mathbf{\mathbf{\Pi}}_{\mathbf{X}_{\omega'}}(\mathbf{x'},t)]=i\hbar\delta(\omega-\omega')\delta^3(\mathbf{x}-\mathbf{x'})\textbf{I} \label{31}
\end{eqnarray}
and
\begin{eqnarray}
[\mathbf{X}_\omega(\mathbf{x},t),\mathbf{X}_{\omega'}(\mathbf{x}',t)]=
[\mathbf{\mathbf{\Pi}}_{\mathbf{X}_{\omega}}(\mathbf{x},t),\mathbf{\mathbf{\Pi}}_{\mathbf{X}_{\omega'}}(\mathbf{x'},t)]=0 \label{32}
\end{eqnarray}
with $\textbf{I}=\mathbf{\hat{x}}\otimes\mathbf{\hat{x}}+\mathbf{\hat{y}}\otimes\mathbf{\hat{y}}+\mathbf{\hat{z}}\otimes\mathbf{\hat{z}}$ the unit dyad. We used the definition $[\mathbf{A}(\mathbf{x}),\mathbf{B}(\mathbf{x}')]=\sum_{\mu,\nu}[A_\mu(\mathbf{x}),B_\nu(\mathbf{x}')]\mathbf{\hat{x}}_\mu\otimes\mathbf{\hat{x}}_\nu$.
This implies the commutation rules:
\begin{eqnarray}
[\mathbf{f}_\omega(\mathbf{x},t),\mathbf{f}^\dagger_{\omega'}(\mathbf{x}',t)]=\delta(\omega-\omega')\delta^3(\mathbf{x}-\mathbf{x'})\textbf{I}. \label{33}
\end{eqnarray} and $[\mathbf{f}_\omega(\mathbf{x},t),\mathbf{f}_{\omega'}(\mathbf{x}',t)]=[\mathbf{f}^\dagger_\omega(\mathbf{x},t),\mathbf{f}^\dagger_{\omega'}(\mathbf{x}',t)]=0$ allowing a clear interpretation of $\mathbf{f}_\omega(\mathbf{x},t)$ and $\mathbf{f}^\dagger_{\omega}(\mathbf{x},t)$ as lowering and rising operators for the bosonic states associated with the matter oscillators.
The quantized Hamiltonian operator is obtained by using the normal-ordered product $H_M\rightarrow:H_M:$ such as 
\begin{eqnarray}
:H_M:=\int d^3\mathbf{x}\int_{0}^{+\infty}d\omega\hbar\omega\mathbf{f}^\dagger_\omega(\mathbf{x},t)\mathbf{f}_\omega(\mathbf{x},t) \label{34}
\end{eqnarray} which allows the elimination of the unphysical vacuum infinite energy.  \\
\indent At that stage it is useful to focus on the polarization density  $\mathbf{P}^{(0)}(\mathbf{x},t)$ which can be equivalently written as
 \begin{eqnarray}
\mathbf{P}^{(0)}(\mathbf{x},t)=\int_0^{+\infty}d\omega\sqrt{\frac{\hbar\sigma_{\omega}(\mathbf{x})}{\pi\omega}}[\mathbf{f}^{(0)}_{\omega}(\mathbf{x},t)\nonumber\\
+\mathbf{f}^{\dagger(0)}_{\omega}(\mathbf{x},t) \label{35}
\end{eqnarray} with by definition $\mathbf{f}^{(0)}_{\omega}(\mathbf{x},t)=\mathbf{f}_{\omega}(\mathbf{x},t_0)e^{-i\omega(t-t_0)}$. This mathematical expression allows for an unambiguous  definition of the Fourier transform $\widetilde{\mathbf{P}}^{(0)}(\mathbf{x},\omega)$ along the real axis which reads:
\begin{eqnarray}
\widetilde{\mathbf{P}}^{(0)}(\mathbf{x},\omega)=\int_0^{+\infty}d\omega'\sqrt{\frac{\hbar\sigma_{\omega'}(\mathbf{x})}{\pi\omega'}}[\mathbf{f}^{(0)}_{\omega'}(\mathbf{x},t_0)\nonumber\\ \cdot e^{i\omega't_0}\delta(\omega-\omega')
+\mathbf{f}^{\dagger(0)}_{\omega'}(\mathbf{x},t_0)e^{-i\omega't_0}\delta(\omega+\omega')]\nonumber\\ \label{36}\end{eqnarray}
The presence of Dirac distributions is key in the reasoning since it allows us to identify the frequency  $\omega'$ in the integral with the pulsation $\omega$ of the Fourier transform. In other words we have $\widetilde{\mathbf{P}}^{(0)}(\mathbf{x},\omega)=\sqrt{\frac{\hbar\sigma_{\omega}(\mathbf{x})}{\pi\omega}}\mathbf{f}^{(0)}_{\omega}(\mathbf{x},t_0)e^{i\omega t_0}$
for $\omega>0$ while 
$\widetilde{\mathbf{P}}^{(0)}(\mathbf{x},\omega)=\sqrt{\frac{\hbar\sigma_{-\omega}(\mathbf{x})}{-\pi\omega}}\mathbf{f}^{\dagger(0)}_{-\omega}(\mathbf{x},t_0)e^{i\omega t_0}$ for $\omega<0$.
Now, Maxwell's equations allow the definition of the current operator $\mathbf{J}^{(0)}(\mathbf{x},t)=\partial_t\mathbf{P}^{(0)}(\mathbf{x},t)$ which implies $\widetilde{\mathbf{J}}^{(0)}(\mathbf{x},\omega)=-i\omega\widetilde{\mathbf{P}}^{(0)}(\mathbf{x},\omega)$. Furthermore we have
\begin{eqnarray}
[\mathbf{f}^{(0)}_\omega(\mathbf{x},t),\mathbf{f}^{\dagger(0)}_{\omega'}(\mathbf{x}',t)]=\delta(\omega-\omega')\delta^3(\mathbf{x}-\mathbf{x'})\textbf{I},\label{37}
\end{eqnarray} and we thus deduce (for $\omega,\omega'\geq 0$ or $\omega,\omega'\leq 0$)
\begin{eqnarray}
[\widetilde{\mathbf{J}}^{(0)}(\mathbf{x},\omega),\widetilde{\mathbf{J}}^{\dagger(0)}(\mathbf{x'},\omega')]\nonumber\\=\omega\frac{\hbar\sigma_{\omega}(\mathbf{x})}{\pi}\delta(\omega-\omega')\delta^3(\mathbf{x}-\mathbf{x'})\textbf{I},
\label{38}
\end{eqnarray} or equivalently $[\widetilde{\mathbf{J}}^{(0)}(\mathbf{x},\omega),\widetilde{\mathbf{J}}^{\dagger(0)}(\mathbf{x'},\omega')]=\langle 0|\widetilde{\mathbf{J}}^{(0)}(\mathbf{x},\omega)\otimes\widetilde{\mathbf{J}}^{\dagger(0)}(\mathbf{x'},\omega')|0\rangle$ (note that Eq.~\ref{38} vanishes if $\omega$ and $\omega'$ have different signs).\\
\indent These formulas are in full agreement with the Gruner-Welsch~\cite{Yeung1996,Gruner1995,Gruner1996} formalism which defines a quantum version of the fluctuation-dissipation theorem for the dielectric medium. We point out that in the Langevin noise approach the fundamental Hamiltonian is given by the fluctuating term
  \begin{eqnarray}
H_M^{(0)}=\int d^3\mathbf{x}\int_{0}^{+\infty}d\omega\hbar\omega\mathbf{f}^{\dagger(0)}_\omega(\mathbf{x},t)\mathbf{f}^{(0)}_\omega(\mathbf{x},t). \label{34fluc}
\end{eqnarray} It was the aim of the original derivation by Huttner and Barnett~\cite{Huttner1992a} (see also ref.~\cite{Suttorp2004b,Philbin2010}) to demonstrate that the Hamiltonian $H_M^{(0)}$ is for all practical needs sufficient for any QED calculations in a dielectric medium. The equivalence will not be studied in this article since it requires a specific study.
 \subsection{Formally quantizing the Maxwell equation}
In order to quantize Maxwell's equations it is here sufficient to solve formally these equations by considering the polarization $\mathbf{P}(\mathbf{x},t)$ as an external source. We will give some details on the derivation here since the use of the $\textbf{F}$ potential is not very common in the context of quantum optics. The strategy will be, like we did for the material field, to start from the Heisenberg  picture in which priority is given to the field evolution equations but with `c numbers' replaced by `q numbers', i.e., operators, which are both time and space dependent.   
First, from Maxwell's equations we  obtain the following second order differential equation
\begin{eqnarray}
\frac{1}{c^2}\partial_t^2\mathbf{F}(\mathbf{x},t)-\boldsymbol{\nabla}^2\mathbf{F}(\mathbf{x},t)-\boldsymbol{\nabla}\times\mathbf{P}(\mathbf{x},t)=0\label{39}
\end{eqnarray} 
which can be  formally solved by using a modal expansion of the potential into plane waves.  For this we write
\begin{eqnarray}
\mathbf{F}(\mathbf{x},t)=\sum_{\alpha,j} q_{\alpha,j}(t)\boldsymbol{\hat{\epsilon}}_{\alpha,j}\Phi_\alpha(\mathbf{x})\label{40}
\end{eqnarray} with $\alpha$ a generic label for  the wave vector $\mathbf{k}_\alpha$,  $\Phi_\alpha(\mathbf{x})=e^{i\mathbf{k}_\alpha\cdot\mathbf{x}}/\sqrt{V}$ (here we consider as it is usually done  the periodical `Box' Born-von Karman expansion in the rectangular box of volume  $V$), $j=1$ or 2, labels the two transverse polarization states with unit vectors $\boldsymbol{\hat{\epsilon}}_{\alpha,1}=\mathbf{k}_\alpha\times\mathbf{\hat{z}}/|\mathbf{k}_\alpha\times\mathbf{\hat{z}}|$, and $\boldsymbol{\hat{\epsilon}}_{\alpha,2}=\hat{\mathbf{k}}_\alpha\times\boldsymbol{\hat{\epsilon}}_{\alpha,1}$ (conventions and more details are given in Appendix B).  The method for solving Maxwell's equations is to transform the second-order differential evolution Eq.~\ref{39} into a set of first-order equations in time. For this we use the variables    
\begin{eqnarray}
c\sqrt{2\hbar \omega_\alpha}c_{\alpha,j}(t)=\frac{d}{dt}q_{\alpha,j}(t)- i\omega_\alpha q_{\alpha,j}(t)\label{44b}\end{eqnarray}
We thus obtain a modal expansion for the fields:
\begin{eqnarray}
\mathbf{F}(\mathbf{x},t)=\sum_{\alpha,j} ic\sqrt{\frac{\hbar}{2\omega_\alpha}}c_{\alpha,j}(t)\boldsymbol{\hat{\epsilon}}_{\alpha,j}\Phi_\alpha(\mathbf{x})+ cc.\nonumber\\
\mathbf{D}(\mathbf{x},t)=\sum_{\alpha,j} -\sqrt{\frac{\hbar \omega_\alpha}{2}}c_{\alpha,j}(t)\hat{\mathbf{k}}_\alpha\times\boldsymbol{\hat{\epsilon}}_{\alpha,j}\Phi_\alpha(\mathbf{x})+ cc.\nonumber\\
\mathbf{B}(\mathbf{x},t)=\sum_{\alpha,j} \sqrt{\frac{\hbar \omega_\alpha}{2}}c_{\alpha,j}(t)\boldsymbol{\hat{\epsilon}}_{\alpha,j}\Phi_\alpha(\mathbf{x})+ cc..\nonumber\\ \label{57b}
\end{eqnarray}
Quantization of those fields holds if we impose the commutation relations:
$[c_{\alpha,j}(t),c_{\beta,k}^{\dag}(t)]=\delta_{\alpha,\beta}\delta_{j,k}$,  
$[c_{\alpha,j}(t),c_{\beta,k}(t)]=0$ and  
$[c_{\alpha,j}^{\dag}(t),c_{\beta,k}^{\dag}(t)]=0$.
We can easily deduce several useful commutation relations  (see Appendix B) like for example
 \begin{eqnarray}
[\mathbf{B}_j(\mathbf{x},t),\mathbf{E}_k(\mathbf{x'},t)]=[\mathbf{B}_j(\mathbf{x},t),\mathbf{D}_k(\mathbf{x'},t)]\nonumber\\=ic\hbar\sum_l\varepsilon_{j,k,l}\partial_l\delta^3(\mathbf{x}-\mathbf{x'})\nonumber\\ \label{56b}
\end{eqnarray} This commutator plays an important role in the Langevin's equation~\cite{Scheel1998} approach and it is here deduced directly from our canonical formalism. Furthermore, using this representation based on the $\mathbf{F}$ potential the Hamiltonian for the pure field $H_F=\int d^3\mathbf{x}(\mathbf{B}^2+\mathbf{D}^2)/2$ becomes:
 \begin{eqnarray}
:H_F(t):=\sum_{\alpha,j}\hbar\omega_\alpha c_{\alpha,j}^\dagger(t)c_{\alpha,j}(t)\label{58}
\end{eqnarray}
which has the usual form for free bosons.\\
\indent We conclude this section by commenting on the use of the electric potential $\mathbf{F}(\mathbf{x},t)$  instead of the more usual magnetic potential $\mathbf{A}(\mathbf{x},t)$. Standard canonical quantization of the electromagnetic field starts from  the separation $\mathbf{B}(\mathbf{x},t)=\boldsymbol{\nabla}\times\mathbf{A}(\mathbf{x},t)$  and $\mathbf{E}(\mathbf{x},t)=\frac{-1}{c}\partial_t\mathbf{A}(\mathbf{x},t)-\boldsymbol{\nabla}V(\mathbf{x},t)$ where $V$ is the scalar potential. The usual standard Lagrangian density reads 
\begin{eqnarray}
\mathcal{L}_s=\frac{\mathbf{E}^2-\mathbf{B}^2}{2}+ \mathbf{A}\cdot\frac{\mathbf{J}}{c} -\rho V+\mathcal{L}_M\label{66}
\end{eqnarray} where $\mathcal{L}_M$ is the same as in Eq.~\ref{3}. This Lagrangian density allows us to derive the same equation of motion as done before. However, since the canonical momentum $\Pi_V$ associated to $V$ is vanishing  we cannot define commutators for those fields and the quantization procedure becomes tricky unless we use the Gupta Bleuer method~\cite{Cohen2}. The usual solution to circumvent this difficulty is to work  exclusively in the Coulomb gauge$\boldsymbol{\nabla}\cdot\mathbf{A}(\mathbf{x},t)=0$ which allows a clear separation between physical and redundant electromagnetic variables (see Ref.~\cite{Cohen2} for a clear analysis of this problem).  The field is thus separated into a transverse contribution $\mathbf{E}_\bot(\mathbf{x},t)=\frac{-1}{c}\partial_t\mathbf{A}(\mathbf{x},t)$, $\mathbf{B}(\mathbf{x},t)=\boldsymbol{\nabla}\times\mathbf{A}(\mathbf{x},t)$, which can be nicely quantized, and into a longitudinal electric field $\mathbf{E}_{||}(\mathbf{x},t)=-\boldsymbol{\nabla}V(\mathbf{x},t)=-\mathbf{P}_{||}(\mathbf{x},t)$ which depends on material fields. Without rewriting here the complete analysis this formalism leads to the following second order equation \begin{eqnarray}
\frac{1}{c^2}\partial_t^2\mathbf{A}(\mathbf{x},t)-\boldsymbol{\nabla}^2\mathbf{A}(\mathbf{x},t)-\frac{\mathbf{J}_\bot(\mathbf{x},t)}{c}=0\label{67}
\end{eqnarray} with $\mathbf{J}(\mathbf{x},t)=\partial_t\mathbf{P}(\mathbf{x},t)$. By using a modal expansion  $\mathbf{A}(\mathbf{x},t)=\sum_{\alpha,j} x_{\alpha,j}(t)\boldsymbol{\hat{\epsilon}}_{\alpha,j}\Phi_\alpha(\mathbf{x})$ and the transformation \begin{eqnarray}
-c\sqrt{2\hbar\omega_\alpha}a_{\alpha,j}(t)=\frac{d}{dt}x_{\alpha,j}(t)- i\omega_\alpha x_{\alpha,j}(t)\label{68}\end{eqnarray}
similar in spirit to Eqs.~44-50  we can rewrite the relevant quantized transverse fields as
\begin{eqnarray}
\mathbf{A}(\mathbf{x},t)=\sum_{\alpha,j} -ic\sqrt{\frac{\hbar}{2\omega_\alpha}}a_{\alpha,j}(t)\boldsymbol{\hat{\epsilon}}_{\alpha,j}\Phi_\alpha(\mathbf{x})+ cc.\nonumber\\
\mathbf{B}(\mathbf{x},t)=\sum_{\alpha,j} \sqrt{\frac{\hbar \omega_\alpha}{2}}a_{\alpha,j}(t)\hat{\mathbf{k}}_\alpha\times\boldsymbol{\hat{\epsilon}}_{\alpha,j}\Phi_\alpha(\mathbf{x})+ cc.\nonumber\\
\mathbf{E}_\bot(\mathbf{x},t)=\sum_{\alpha,j} \sqrt{\frac{\hbar \omega_\alpha}{2}}a_{\alpha,j}(t)\boldsymbol{\hat{\epsilon}}_{\alpha,j}\Phi_\alpha(\mathbf{x})+ cc.\nonumber\\ \label{69}
\end{eqnarray} which must be compared with Eq.~\ref{57b}. The most important difference between the formalism using $\textbf{A}$ and the one based on $\textbf{F}$ is that the later use only local and causal electromagnetic properties such as $D$, $B$, and $P$ while the former use a separation between transverse and longitudinal fields and currents which not are causal  when taken separately~\cite{Cohen2}. It is therefore an advantage of our method to eliminate  such unphysical separation from the ground. \\
\indent Quantization can be easily done by imposing the commutation rules $[a_{\alpha,j}(t),a_{\beta,k}^{\dag}(t)]=\delta_{\alpha,\beta}\delta_{j,k}$
and $[a_{\alpha,j}(t),a_{\beta,k}(t)]=[a_{\alpha,j}^{\dag}(t),a_{\beta,k}^{\dag}(t)]=0$ from which we deduce the same field commutators as the one discussed previously and derived in the appendix  (in particular Eq.~\ref{56b}) .\\
\indent The canonical Lagrangian density \begin{eqnarray}\mathcal{L}_c=\frac{\mathbf{E}_\bot^2-\mathbf{B}^2}{2}+ \mathbf{A}\cdot\frac{\mathbf{J}_\bot}{c} -\rho V/2+\mathcal{L}_M\label{canonlag}\end{eqnarray}
with $\rho=-\boldsymbol{\nabla}\cdot\mathbf{P}=-\boldsymbol{\nabla}\cdot\mathbf{P}_{||}$  (implying $\int d^3\mathbf{x}\mathbf{E}_{||}^2=\int d^3\mathbf{x}\mathbf{P}_{||}^2=\int d^3\mathbf{x}\rho V$) allows us to introduce the canonical Hamiltonian:
\begin{eqnarray}
H_c(t)=\int d^3\mathbf{x}[\frac{\mathbf{E}_\bot^2+\mathbf{B}^2}{2}-\mathbf{A}\cdot\frac{\mathbf{J}_\bot}{c} +\frac{\mathbf{E}_{||}^2}{2}]+H_{c,M},\nonumber\\
\label{70}
\end{eqnarray}
where the canonical Hamiltonian  for the matter field differs from $H_M$, as given in Eq.~12, by the amount  $H_{c,M}-H_M=\mathbf{A}\cdot\frac{\mathbf{J}_\bot}{c}$. This results from a different canonical momentum for the matter field $\mathbf{\mathbf{\Pi}}_\mathbf{c,X_\omega}=\partial_t\mathbf{X}_\omega+\sqrt{\frac{2\sigma_\omega(\mathbf{x})}{\pi}}\mathbf{A}$. 
The canonical Hamiltonian $H_c(t)$ can be also expressed as a functional of the canonical variables 
and we get 
\begin{eqnarray}
H(t)=\int d^3\mathbf{x}[\frac{c^2\mathbf{\mathbf{\Pi}}_\mathbf{A}^2+\boldsymbol{\nabla}\times\mathbf{A}^2}{2} +\frac{\mathbf{P}_{||}^2}{2}]\nonumber\\
+\int d^3\mathbf{x}\int_{0}^{+\infty}d\omega\frac{(\mathbf{\mathbf{\Pi}}_\mathbf{c,X_\omega}-\sqrt{\frac{2\sigma_\omega(\mathbf{x})}{\pi}}\mathbf{A})^2+\omega^2\mathbf{X}_\omega^2}{2}
\nonumber\\ \label{10c}
\end{eqnarray} which should be compared with Eq.~\ref{10b}.\\
Despite these differences the terms proportional to $\mathbf{A}\cdot\frac{\mathbf{J}_\bot}{c}$  cancel out in Eq.~\ref{70} and we get the remarkable result
\begin{eqnarray}
H_c(t)=\int d^3\mathbf{x}\frac{\mathbf{B}^2+\mathbf{E}^2}{2}+H_M(t)=H(t)\label{71}
\end{eqnarray} with $\int d^3\mathbf{x}\frac{:\mathbf{B}^2+\mathbf{E}_\bot^2:}{2}=\sum_{\alpha,j}\hbar\omega_\alpha a_{\alpha,j}^\dagger(t)a_{\alpha,j}(t)$. Eq.~\ref{71} shows that the two formalisms based on $\mathbf{A}$ or  $\mathbf{F}$ should be equivalent (this is also confirmed by the fact that we obtain the same commutators for the electromagnetic fields in both formalisms). Rigorously the correspondence between these two languages can be done by equating the electromagnetic fields obtained with both methods. We thus obtain a relationship between the $x_{\alpha,j}(t)$  and $q_{\alpha,j}(t)$ variables.  This is done easily by remarking that we have $\mathbf{D}=\mathbf{E}+\mathbf{P}=\mathbf{E}_\bot+\mathbf{P}_\bot$.  As explained in Appendix C we get 
\begin{eqnarray}
c_{\alpha,1}(t)= -a_{\alpha,2}(t)-\frac{1}{\sqrt{2\hbar\omega_\alpha}}P_{\alpha,2}(t)\nonumber\\
c_{\alpha,2}(t)= a_{\alpha,1}(t)+\frac{1}{\sqrt{2\hbar\omega_\alpha}}P_{\alpha,1}(t).\label{76}
\end{eqnarray} with the definition \begin{eqnarray} P_{\alpha,j}(t)=\int d^{3}\mathbf{x}\int_{0}^{+\infty}d\omega\sqrt{\frac{2\sigma_\omega(\mathbf{x})}{\pi}}\mathbf{X}_{\omega}(\mathbf{x},t)\nonumber\\ \cdot\boldsymbol{\hat{\epsilon}}_{\alpha,j}\Phi_\alpha^\ast(\mathbf{x})\label{75}\end{eqnarray}
We emphasize that the transformation between two formalisms based on $\textbf{F}$ and $\textbf{A}$ potentials can be handled differently using the so called Power-Zienau unitary transformation~\cite{Cohen2}. In this article we did not use this approach (see however Appendix C) and we instead introduced the dual Lagrangian given in Eq.~\ref{1}. We will however discuss further in the conclusion the relationship between the different formalisms.\\   
\indent Before leaving this section we emphasize that all the dynamical equations used previously could be equivalently obtained from the Heisenberg equation, which, for an operator $A(t)$, reads as   $i\hbar\frac{d}{dt}A(t)=[A(t),H(t)]$ where $H(t)$ is the full Hamiltonian operator.  This was checked for the equations used in this work. However, we point out that the Hamiltonian formalism has some intrinsic limitations since it relies on some convergence hypothesis which were briefly mentioned in the Section II. Indeed, since for fields we discuss the problem of radiating system the conservation of the energy in a fixed volume is in general not valid unless we accept some specific boundary conditions at spatial infinity  (i.e., in general the Poynting vector flow at infinity does not vanish). At the opposite, we could take the quantized dynamical equations as fundamental postulates of the theory without relying on the Hamilton operator and checking the consistency  of the formulas  at hand (see also Ref.~\cite{Sipe1995} for a wave-function analysis of the photon dynamics in vacuum).  However, both approaches give of course similar results as far as the boundary conditions are taken into considerations.

\section{Comparison with the minimal coupling and multipolar representation}
While the previous analysis was given in order to model macroscopic quantum electrodynamics in dielectric media, it is particularly important to discuss the microscopic physical origin of the model in order to evaluate the hypothesis, limitations  and possible generalization of the approach. 
First, observe as a reminder that the minimal coupling scheme associated with the Lagrangian density Eq.~\ref{canonlag} corresponds actually to the continuous limit of the non-relativistic Lagrange function~\cite{Cohen2}:
 \begin{eqnarray}L_c=\int d^3\mathbf{x}\frac{\mathbf{E}_\bot^2-\mathbf{B}^2}{2}+ \mathbf{A}\cdot\frac{\mathbf{J}_\bot}{c} -\frac{\rho V}{2}+\sum_n T_n- U\label{canonlag2}\nonumber\\
\end{eqnarray} 
where $\sum_n T_n$ is the discrete sum  of the kinetic energy terms $T_n=\frac{1}{2}m_n(\frac{d\textbf{x}_n(t)}{dt})^2$ associated with the point-like particle with individual mass $m_n$ and spatial coordinates $\textbf{x}_n(t)$ ($n=1,...N$ is an integer labeling the particles)~\cite{Cohen2}. We also included an interaction potential $U(\textbf{x}_1,...,\textbf{x}_N)$ which depends on the  $N$ particle coordinates. In this description the symmetrized electric current is 
\begin{eqnarray}
\textbf{J}=\frac{1}{2}\sum_n e_n\frac{d\textbf{x}_n(t)}{dt}\delta^3(\textbf{x}-\textbf{x}_n)\nonumber\\+\frac{1}{2}\sum_n e_n\delta^3(\textbf{x}-\textbf{x}_n)\frac{d\textbf{x}_n(t)}{dt}\label{symmet}\end{eqnarray}  (the symmetrization is necessary for satisfying the commutation relations) while the charge density is $\rho=\sum_n e_n\delta^3(\textbf{x}-\textbf{x}_n)$  with $e_n$ the individual charge of each moving individual electrons. For neutral matter this total charge  $\sum_n e_n$ is of course  globally neutralized by the static charges of the atomic nuclei. Moreover, in the coulomb Gauge description the fundamental current is the transverse current $\textbf{J}_\bot$   where  the transverse delta function $\boldsymbol{\delta_{\bot}}(\textbf{x}-\textbf{x}_n)$, defined in Eq.~\ref{53}, replaces $\delta^3(\textbf{x}-\textbf{x}_n)$.  From Eq.~\ref{canonlag} we deduce the direct generalization of  the Hamiltonian $H_c$ given by Eq.~\ref{70}
\begin{eqnarray}
H_c(t)=\int d^3\mathbf{x}[\frac{\mathbf{E}_\bot^2+\mathbf{B}^2}{2}-\mathbf{A}\cdot\frac{\mathbf{J}_\bot}{c} +\frac{\mathbf{E}_{||}^2}{2}]+\sum_n T_n+U\nonumber\\
\label{70new}
\end{eqnarray} 
with $T_n=\frac{(\textbf{p}_n-e_n\textbf{A}(\textbf{x}_n,t)/c)^2}{2m_n}$ is explicitly written using the particle canonical momentum $\textbf{p}_n=m_n \frac{d\textbf{x}_n(t)}{dt}+ e_n\textbf{A}(\textbf{x}_n,t)/c$ associated with $\textbf{x}_n$. Moreover, the full evolution leads straightforwardly to Maxwell's equations~\cite{Cohen2} and to the quantized Lorentz force dynamical equation:
\begin{eqnarray}
m_n\frac{d^2\textbf{x}_n(t)}{dt^2}=e_n\textbf{E}(\textbf{x}_n,t)-\frac{\partial}{\partial \textbf{x}_{n}}U\nonumber\\+e_n\frac{1}{2c}\frac{d\textbf{x}_n(t)}{dt}\times\textbf{B}(\textbf{x}_n,t)
-e_n\frac{1}{2c}\textbf{B}(\textbf{x}_n,t)\times\frac{d\textbf{x}_n(t)}{dt}
\end{eqnarray}
where the symmetrization is required from the ground.\\
\indent The usual approximation made to model a dielectric medium is to expand  the Lorentz force as a power of the relative coordinate $\boldsymbol{\xi}_n$ between the $n\textbf{}^{th}$ electron and the associated nuclei (we suppose only one electron per atom). The classical result \cite{Novotny} implies the standard dipolar approximation leading to the force
 \begin{eqnarray}
M_n\frac{d^2\textbf{X}_{n}(t)}{dt^2}\simeq e_n\sum_{i=1}^{i=3}\xi_{n,i}\frac{\partial}{\partial \textbf{X}_{n}}E_i(\textbf{X}_n,t)\nonumber\\
+e_n\frac{\partial}{\partial t}[\boldsymbol{\xi}_n\times\frac{\textbf{B}(\textbf{X}_n,t)}{c}]\label{retrucu}
\end{eqnarray}
 where $\textbf{X}_n$ is the center-of-mass coordinate of the electron-nucleus system with total mass $M_n$. In many applications, e.g., with harmonic excitation,  the time derivative in the second terms average to zero so that only the gradient force survives~\cite{Novotny}. This center-of-mass equation is not exhausting the dynamics of the dielectric system and for optical application the fundamental relation is the internal dynamics which is given by the equation
 \begin{eqnarray}
\mu_n\frac{d^2\boldsymbol{\xi}_{n}(t)}{dt^2}\simeq e_n \textbf{E}(\textbf{X}_n,t)-\frac{\partial}{\partial \boldsymbol{\xi}_{n}}U_n(\boldsymbol{\xi}_{n})\nonumber\\
+e_n\frac{1}{2c}\frac{d\textbf{X}_n(t)}{dt}\times\textbf{B}(\textbf{X}_n,t)
-e_n\frac{1}{2c}\textbf{B}(\textbf{X}_n,t)\times\frac{d\textbf{X}_n(t)}{dt} \nonumber\\
\label{internal}
\end{eqnarray}
where we have made the assumption $U=\sum_n U_n(\boldsymbol{\xi}_{n})$ and introduced the reduced mass $\mu_n=m^{(e)}_n m^{(n)}_n/M_n$ of the electron-nucleus pair. Clearly, Eq.~\ref{8} is a special case of this internal dynamics corresponding to an harmonic interaction potential $U_n$ and to the static condition $\frac{d\textbf{X}_n(t)}{dt}=0$  removing the magnetic  Lorentz force.  We now point out that while the minimal coupling description is clearly sufficient for many purposes a different but rigorously equivalent way to describe the electromagnetic coupling in dielectric systems is to use the multipolar representation~\cite{Cohen2,Zienau,Craig,Sipe1995}. It  is obtained by adding a term $-\frac{d}{dt}[\int d^3\mathbf{x}\frac{\mathbf{A}\cdot\mathbf{P}}{c}]$ to the canonical Lagrangian given by Eq.~\ref{canonlag2}. Through this canonical transformation and after introducing the density of electric and magnetic polarization $\textbf{P}$ and $\textbf{M}$ (which for neutral systems are connected to $\textbf{J}$ and $\rho$ by $\textbf{J}=\partial_t \textbf{P}+ c\boldsymbol{\nabla}\times \mathbf{M}$, and $\rho=-\boldsymbol{\nabla}\times \mathbf{P}$) we obtain the Lagrange function: 
\begin{eqnarray}L_{\textrm{multi.}}=\int d^3\mathbf{x}[\frac{\mathbf{D}^2-\mathbf{B}^2-\mathbf{P}^2}{2}+ \mathbf{M}\cdot\mathbf{B}] \nonumber\\ +\sum_n T_n-U\label{multipolag}.\end{eqnarray}
 With this definition we deduce the canonical momenta $\mathbf{\mathbf{\Pi}}_{\mathbf{A},\textrm{multi.}}=-\frac{\mathbf{D}}{c}$, and $\textbf{p}_{n,\textrm{multi.}}=m_n \frac{d\textbf{x}_n(t)}{dt}+\frac{\partial}{\partial \dot{\textbf{x}}_n}[\int d^3\mathbf{x}\mathbf{M}\cdot\mathbf{B}] $ (which differ from the usual minimal coupling values) and a different Hamiltonian
 \begin{eqnarray}H_{\textrm{multi.}}=\int d^3\mathbf{x}[\frac{\mathbf{D}^2+\mathbf{B}^2+\mathbf{P}^2}{2}-\mathbf{M}\cdot\mathbf{B}-\mathbf{P}\cdot\mathbf{D}] \nonumber\\
+\sum_n T_n+U+\sum_n\dot{\textbf{x}}_n\cdot\frac{\partial}{\partial \dot{\textbf{x}}_n}[\int d^3\mathbf{x}\mathbf{M}\cdot\mathbf{B}]\label{multipoham}.\end{eqnarray}
These  equations are sufficient to obtain directly the macroscopic Maxwell equations for any dielectric and magnetic media. In the general case the electric dipole density $\textbf{P}$ is expressed in the classical problem by a line integral as \cite{Cohen2,Craig,Novotny} $\textbf{P}(\textbf{x},t)=\sum_n e_n \boldsymbol{\xi}_n(t)\int_0^1 du\delta^3(\textbf{x}-u\boldsymbol{\xi}_n-\textbf{x}^{(N)}_n)$  with $\textbf{x}^{(N)}_n$ the coordinate of the  $n^{\textrm{th}}$ nuclei of electric charge $-e_n$ and $\boldsymbol{\xi}_n+\textbf{x}^{(N)}_n:=\textbf{x}^{(e)}_n$ is the coordinate of the electron of electric charge $e_n$. We have generally 
\begin{eqnarray}\rho=-\boldsymbol{\nabla}\cdot \mathbf{P}=\sum e_n[\delta^3(\textbf{x}-\textbf{x}^{(e)}_n)-\delta^3(\textbf{x}-\textbf{x}^{(N)}_n)]\end{eqnarray} in order to preserve the total charge cancellation of the dielectric medium.  We obtain similarly 
$\textbf{M}(\textbf{x},t)=\sum_n \frac{e_n}{c} \int_0^1 du \delta^3(\textbf{x}-u\boldsymbol{\xi}_n-\textbf{x}^{(N)}_n)\boldsymbol{\xi}_n(t)\times [u \frac{d\boldsymbol{\xi}_n}{dt}+\frac{d\textbf{x}^{(N)}_n}{dt}]$ and thus the classical (non symmetric) current 
\begin{eqnarray} \textbf{J}= \sum e_n[\frac{d\textbf{x}^{(e)}_n}{dt}\delta^3(\textbf{x}-\textbf{x}^{(e)}_n)-\frac{d\textbf{x}^{(N)}_n}{dt}\delta^3(\textbf{x}-\textbf{x}^{(N)}_n)].\end{eqnarray}  We point out that the rigorous extension of the line integral formulas from classical to quantum regime is not a trivial task since we should consider carefully the non commutation  of conjugate variables,  i.e., positions and velocities~\cite{Raab,Wooley}. It is alternatively possible to use the second quantized formalism involving the wave function operators $\Psi(\mathbf{x},t)$, $\Psi^\dagger(\mathbf{x},t)$ associated with particles  to remove ambiguities~\cite{Power1983}. Hence, at the end we will obtain a symmetrized current like in Eq.~\ref{symmet}.  With these definitions it is not difficult to obtain $H_{\textrm{multi.}}=\int d^3\mathbf{x}[\frac{\mathbf{E}^2+\mathbf{B}^2}{2}] +\sum_n T_n+U$ allowing us to identify the multipolar Hamiltonian with the total energy. For the present purpose we are interested in the crude dipolar approximation $\textbf{M}(\textbf{x},t)\approx 0$, $\textbf{P}(\textbf{x},t)\approx\sum_n e_n \boldsymbol{\xi}_n(t)\delta^3(\textbf{x}-\textbf{X}_n)$ with $\textbf{X}_n\simeq\textbf{x}^{(N)}_n$  the center-of-mass coordinate of the $n^{\textrm{th}}$ electron-nucleus pair (in order to remove the singularity of the delta function $\delta^3(\textbf{x})$ and the infinite self-interaction coming  from the term $\textbf{P}^2/2$ in the Lagrangian and Hamiltonian one can alternatively introduce a narrow function $\Delta(\textbf{x})$~\cite{Jackson1999}peaked on the origin). Within this approximation the total force acting on the electron-nucleus system leads to the dynamical law:
\begin{eqnarray}
M_n\frac{d^2\textbf{X}_{n}(t)}{dt^2}\simeq e_n\sum_{i=1}^{i=3}\xi_{n,i}\frac{\partial}{\partial \textbf{X}_{n}}E_i(\textbf{X}_n,t)\label{crudu}
\end{eqnarray} without the time derivative term in Eq.~\ref{retrucu} (Eq.~\ref{retrucu} can be justified by relaxing the approximation done on $\textbf{P}$ and $\textbf{M}$~\cite{Wubs2003,Lembessis1993}, i.e. by introducing a Rontgen current in the Hamiltonian~\cite{Wilkens1994}). These forces can be summed over the different particles and leads to the total force acting on the medium: $\textbf{F}_{\textrm{total}}=\int d^3\mathbf{x}\sum_{i=1}^{i=3}P_i(\textbf{x},t)\boldsymbol{\nabla}E_i(\textbf{x},t)$ (the inclusion of  internal forces between particles $n$ and $m$ will not change the final result as far as these additional terms cancel out over the summation and integration). Importantly, in the multipolar framework the internal dynamical motion given by Eq.~\ref{internal} is kept unchanged, meaning that all optical QED applications in a dielectric medium can be handled equivalently  with the minimal coupling or multipolar scheme. \\
\indent What is however key for the present article is to check the self-consistency of our dual representation discussed in previous sections and based on the electric potential $\textbf{F}$. To do this, observe that the obvious generalization of Lagrangian Eq.~\ref{1} reads:
  \begin{eqnarray}
L=\int d^3\mathbf{x}[\frac{\mathbf{B}^2-\mathbf{D}^2}{2}+ \mathbf{F}\cdot\boldsymbol{\nabla}\times\mathbf{P} -\frac{\mathbf{P}^2}{2}]+\sum_n T_n-U_n\nonumber\\ \label{gener}
\end{eqnarray} As explained in Appendix A this Lagrangian is actually obtained  from the multipolar Lagrangian  Eq.~\ref{multipolag} by using some duality relations between electric and magnetic quantities. By using  the electromagnetic dual-Lagrangian $L_{m,\textrm{multi.}}$ (see Eq.~\ref{multibiss}) with $\textbf{M=}0$ we have  $L=L_{m,\textrm{multi.}}+\sum_n T_n-U_n-\int d^3\mathbf{x}\frac{\mathbf{P}^2}{2}$. The added terms only depend on the particle variables and are not affecting the Maxwell equations. The inclusion of the supplementary potential $\int d^3\mathbf{x}\frac{\mathbf{P}^2}{2}$ is necessary in order to justify the equation of motion Eqs.~\ref{internal} and \ref{crudu}. As explained before the inclusion of internal forces  leads to Eq.~\ref{internal} and this is true in this representation as well.  In this way the Huttner-Barnett Lagrangian $L_M$ (see Eq.~\ref{2}) is only a particular case leading to Eq.~\ref{8}. Still, in all cases we can easily obtain the macroscopic Maxwell equations as well as the dynamical motion laws for the particles, i.e., Eq.~\ref{crudu} without the time derivative term. Clearly this means  that the approach outlined in this paper is equivalent to the minimal coupling or multipolar Lagrangian within the crude dipolar approximation, which is what we wanted to demonstrate (more on this and on the duality relation is given in Appendix A).\\
\section{Conclusion}
To conclude,  we introduced a dual representation of the quantized electromagnetic field in dielectric media based on the transverse  electric potential vector  $\textbf{F}$ instead of the usual magnetic potential vector $\textbf{A}$ in the Coulomb gauge. The method, contrarily to the usual minimal coupling representation, involves only physical, i.e., local and causal quantities.    We showed that our approach is well adapted to the analysis of neutral systems without magnetic property. The dual Lagrangian formalism  is equivalent to both standard multipolar and minimal coupling representations within the crude dipolar approximation. This allowed us to re-derive the standard QED equations within the Huttner-Barnett model for quantized polaritons~\cite{Huttner1992a,Suttorp2004b,Philbin2010}.   The equivalence with the usual commutation relations given in the Langevin noise approach~\cite{Gruner1996,Scheelreview2008}  was discussed and the Hamiltonian formalism was justified from the ground within our dual approach.
In future works we plan to apply the dual Lagrangian formalism to general QED questions in homogeneous and inhomogeneous dielectric media involving coupling with quantum emitters. We believe our work will motivate further studies concerning duality relations and quantization in dense optical media. 
  \section{Acknowledgments}
This work was supported by Agence Nationale de la Recherche (ANR), France,
through the SINPHONIE (ANR-12-NANO-0019) and PLACORE (ANR-13-BS10-0007) grants. The author gratefully acknowledge stimulating discussions with G. Bachelier, S. Huant, and C. Genet.  
\appendix
\section{Duality in dielectric and magnetic media}
\indent It is interesting to understand that the motivation for the  Lagrangian given by Eqs.~\ref{1} and \ref{gener} is connected to a particular form of duality existing for Maxwell's equations. Consider indeed the most general set of Maxwell's  equations for continuous media:  
\begin{eqnarray}
\boldsymbol{\nabla}\times\mathbf{B}=\frac{1}{c}\partial_t\mathbf{E}+\frac{\mathbf{J}_e}{c}, &\boldsymbol{\nabla}\cdot\mathbf{B}=0\nonumber\\
\boldsymbol{\nabla}\times\mathbf{E}=-\frac{1}{c}\partial_t\mathbf{B}, &\boldsymbol{\nabla}\cdot\mathbf{E}=\rho_e \label{maxou}
\end{eqnarray}
where the electric current and charge density are given by $\textbf{J}_e=\partial_t \textbf{P}+ c\boldsymbol{\nabla}\times \mathbf{M}$, and $\rho_e=-\boldsymbol{\nabla}\cdot \mathbf{P}$.  By introducing $\textbf{H}=\textbf{B}-\textbf{M}$ and $\textbf{D}=\textbf{E}+\textbf{P}$  Eqs.~\ref{maxou} transform as:
 \begin{eqnarray}
\boldsymbol{\nabla}\times\mathbf{D}=\frac{-1}{c}\partial_t\mathbf{H}+\frac{\mathbf{J}_m}{c}, &\boldsymbol{\nabla}\cdot\mathbf{D}=0\nonumber\\
\boldsymbol{\nabla}\times\mathbf{H}=\frac{1}{c}\partial_t\mathbf{D}, &\boldsymbol{\nabla}\cdot\mathbf{H}=\rho_m \label{maxoub}
\end{eqnarray}   with the magnetic current and charge density given by:  $\textbf{J}_m=\partial_t \textbf{M}- c\boldsymbol{\nabla}\times \mathbf{M}$, and $\rho_m=-\boldsymbol{\nabla}\cdot \mathbf{M}$.  There is clearly a duality relation between Eqs.~\ref{maxou} and \ref{maxoub} and we go from the first to the second by the  dual replacement:
 \begin{eqnarray}
	\mathbf{E}\rightarrow -\mathbf{H},& 	\mathbf{B}\rightarrow \mathbf{D}\nonumber\\
	\mathbf{J}_e\rightarrow -\mathbf{J}_m,& 	\mathbf{P}\rightarrow -\mathbf{M}\nonumber\\
		\rho_e\rightarrow -\rho_m,& 	\mathbf{M}\rightarrow \mathbf{P}\nonumber\\
\mathbf{A}\rightarrow \mathbf{F},& 	V\rightarrow V'
\end{eqnarray} where by definition $\textbf{H}= \frac{1}{c}\partial_t\mathbf{F}+\boldsymbol{\nabla}V'$ which depends on the dual potentials $\mathbf{F}$ and $V'$. Now,  Eqs.~\ref{maxou} can be derived  from the standard electromagnetic canonical Lagrangian density $\mathcal{L}_e=\frac{\mathbf{E}^2-\mathbf{B}^2}{2}+ \mathbf{A}\cdot\frac{\mathbf{J}_e}{c} -\rho_e V$, i.e., Eq.~\ref{66} without the material part.  This means that  the dual set  of equations Eqs.~\ref{maxoub} can equivalently be derived from the `standard' Lagrange function: 
\begin{eqnarray}
\mathcal{L}_{m,s}=\frac{\mathbf{H}^2-\mathbf{D}^2}{2} - \mathbf{F}\cdot\frac{\mathbf{J}_m}{c} +\rho_m. V' \label{66new}
\end{eqnarray}  The multipolar analog of Eq.~\ref{multipolag} $L_{m,\textrm{multi.}}=L_{m,s}+\frac{d}{dt}[\int d^3\mathbf{x}\frac{\mathbf{F}\cdot\mathbf{M}}{c}]$ 
reads in this language:
\begin{eqnarray}
L_{m,\textrm{multi.}}=\int d^3\mathbf{x}[\frac{\mathbf{B}^2-\mathbf{D}^2-\mathbf{M}^2}{2}+ \mathbf{P}\cdot\mathbf{D}]\label{multibiss}\end{eqnarray}
 and the Hamiltonian becomes 
\begin{eqnarray}
H_{m}=\int d^3\mathbf{x}[\frac{\mathbf{B}^2+\mathbf{D}^2+\mathbf{M}^2}{2}- \mathbf{P}\cdot\mathbf{D}-\mathbf{B}\cdot\mathbf{M}]\end{eqnarray}
This duality is actually reminiscent of the early work made in the $19^{th}$ century when Coulomb and Biot proposed an interpretation of magnetism in term of magnetic charge (in analogy with electrostatic ) while Ampere proposed to give a electric current origin to magnetism~\cite{Darrigol}. The equivalence is however not complete since there is no magnetic monopole in the standard electromagnetism approach. More precisely if we consider the total force acting on a medium characterized by $\textbf{P}$ and $\textbf{M}$ in the standard Maxwell representation Eqs.~\ref{maxou} we get from Noether's theorem:
  \begin{eqnarray}
	\textbf{F}_{e,\textrm{total}}=\int d^3\mathbf{x}[\rho_e\textbf{E}+\frac{\textbf{J}_e\times\textbf{B}}{c}]=\frac{d}{dt}[\int d^3\mathbf{x}\frac{\mathbf{P}\times\mathbf{B}}{c}]\nonumber\\
	+\int d^3\mathbf{x}\sum_{i=1}^{i=3}[P_i(\textbf{x},t)\boldsymbol{\nabla}E_i(\textbf{x},t)+M_i(\textbf{x},t)\boldsymbol{\nabla}B_i(\textbf{x},t)] \nonumber\\
	\end{eqnarray} while the dual representation Eqs.~\ref{maxoub} leads to the total force
	\begin{eqnarray}
	\textbf{F}_{m,\textrm{total}}=\int d^3\mathbf{x}[\rho_m\textbf{H}-\frac{\textbf{J}_m\times\textbf{D}}{c}]=\frac{d}{dt}[\int d^3\mathbf{x}\frac{\mathbf{D}\times\mathbf{M}}{c}]\nonumber\\
	+\int d^3\mathbf{x}\sum_{i=1}^{i=3}[P_i(\textbf{x},t)\boldsymbol{\nabla}E_i(\textbf{x},t)+M_i(\textbf{x},t)\boldsymbol{\nabla}B_i(\textbf{x},t)].\nonumber\\ 
	\end{eqnarray} Clearly in general $\textbf{F}_{e,\textrm{total}}$ differs from $\textbf{F}_{m,\textrm{total}}$ unless the time derivatives cancel. This special case occurs in the magnetostatic limit but also for oscillating motions when time average removes the time derivative terms as in Eq.~\ref{crudu} (this is the case for fluctuating forces considered in nano-photonics~\cite{Novotny})  
\section{Plane waves modal expansion  and quantization of the electromagnetic field }
In order to quantize the electromagnetic field using a plane-wave modal expansion we first remind several mathematical properties of the Born von Karman expansion method. First, we have  the normalization (obtained from $\boldsymbol{\nabla}^2\Phi_\alpha(\mathbf{x})+k_\alpha^2\Phi_\alpha(\mathbf{x})=0$, with $k_\alpha=|\mathbf{k}_\alpha|=\omega_\alpha/c$):
\begin{eqnarray}
\int_Vd^3\mathbf{x}\Phi_\alpha(\mathbf{x})\Phi_\beta^\ast(\mathbf{x})=\delta_{\alpha,\beta}.\label{41}
\end{eqnarray}
Second, we have the following symmetries for $\mathbf{k}_\alpha=-\mathbf{k}_{-\alpha}$:
\begin{eqnarray}
\Phi_{-\alpha}(\mathbf{x})=\Phi_\alpha^\ast(\mathbf{x})\nonumber\\
\boldsymbol{\hat{\epsilon}}_{-\alpha,1}=-\boldsymbol{\hat{\epsilon}}_{\alpha,1}, \boldsymbol{\hat{\epsilon}}_{-\alpha,2}=+\boldsymbol{\hat{\epsilon}}_{\alpha,2}.\label{42}
\end{eqnarray} which we summarize as $\boldsymbol{\hat{\epsilon}}_{-\alpha,j}=\eta_{j}\boldsymbol{\hat{\epsilon}}_{\alpha,j}$ with $\eta_1=-1$ and $\eta_2=+1$.
The field being real valued, i.e., $\mathbf{F}(\mathbf{x},t)=\mathbf{F}^\ast(\mathbf{x},t)$ we deduce
\begin{eqnarray}
q_{-\alpha,j}(t)=\eta_j q_{\alpha,j}^\ast(t)\label{43}
\end{eqnarray}
In order to solve Eq.~39 it is useful to introduce the  auxiliary variables:
\begin{eqnarray}
Z^{(\pm)}_{\alpha,j}(t)=\frac{d}{dt}q_{\alpha,j}(t)\pm i\omega_\alpha q_{\alpha,j}(t)\label{44}\end{eqnarray}
i.e.,\begin{eqnarray}
q_{\alpha,j}(t)=\frac{Z^{(+)}_{\alpha,j}(t)-Z^{(-)}_{\alpha,j}(t)}{2i\omega_\alpha}\nonumber\\
\frac{d}{dt}q_{\alpha,j}(t)=\frac{Z^{(+)}_{\alpha,j}(t)+Z^{(-)}_{\alpha,j}(t)}{2}.\label{45}
\end{eqnarray}
From the reality requirement Eq.~\ref{43} we deduce:
\begin{eqnarray}
Z^{(\pm)}_{-\alpha,j}(t)=\eta_j{Z^{(\mp)}_{\alpha,j}}^\ast(t).\label{46}
\end{eqnarray}
These relations and definitions lead to:
\begin{eqnarray}
\mathbf{F}(\mathbf{x},t)=\sum_{\alpha,j} \frac{Z^{(-)}_{\alpha,j}(t)}{-2i\omega_\alpha}\boldsymbol{\hat{\epsilon}}_{\alpha,j}\Phi_\alpha(\mathbf{x})+ cc.\nonumber\\
\mathbf{D}(\mathbf{x},t)=\sum_{\alpha,j} \frac{-Z^{(-)}_{\alpha,j}(t)}{2c}\hat{\mathbf{k}}_\alpha\times\boldsymbol{\hat{\epsilon}}_{\alpha,j}\Phi_\alpha(\mathbf{x})+ cc.\nonumber\\
\mathbf{B}(\mathbf{x},t)=\sum_{\alpha,j} \frac{Z^{(-)}_{\alpha,j}(t)}{2c}\boldsymbol{\hat{\epsilon}}_{\alpha,j}\Phi_\alpha(\mathbf{x})+ cc..\nonumber\\ \label{47}
\end{eqnarray}
In this  representation the field equations  read
\begin{eqnarray}
\dot{Z}^{(\pm)}_{\alpha,j}(t)=\pm i\omega_\alpha Z^{(\pm)}_{\alpha,j}(t) +S_{\alpha,j}(t)\label{59}
\end{eqnarray} with the source term
\begin{eqnarray}
S_{\alpha,j}(t)=c^2\int d^3\mathbf{x}\boldsymbol{\nabla}\times\mathbf{P}(\mathbf{x},t)\cdot\boldsymbol{\hat{\epsilon}}_{\alpha,j}\Phi_\alpha^\ast(\mathbf{x})\label{60}
\end{eqnarray}
One strategy for quantizing the electromagnetic field is to introduce   rizing $c_{\alpha,j}^\dagger(t)$ and lowering  $c_{\alpha,j}(t)$ photon operators defined by 
\begin{eqnarray}
\frac{Z^{(-)}_{\alpha,j}(t)}{2c}=\sqrt{\frac{\hbar \omega_\alpha}{2}}c_{\alpha,j}(t)\label{48}
\end{eqnarray}
with the commutators $[c_{\alpha,j}(t),c_{\beta,k}^{\dag}(t)]=\delta_{\alpha,\beta}\delta_{j,k}$ and $[c_{\alpha,j}(t),c_{\beta,k}(t)]=[c_{\alpha,j}^{\dag}(t),c_{\beta,k}^{\dag}(t)]=0$. We thus have
\begin{eqnarray}
q_{\alpha,j}(t)=c\sqrt{(2\hbar\omega_\alpha)}\frac{\eta_j c^\dagger_{-\alpha,j}(t)-c_{\alpha,j}(t)}{2i\omega_\alpha}\nonumber\\
\frac{d}{dt}q_{\alpha,j}(t)=c\sqrt{(2\hbar\omega_\alpha)}\frac{\eta_j c^\dagger_{-\alpha,j}(t)+c_{\alpha,j}(t)}{2}\label{49}
\end{eqnarray}
This  means: 
\begin{eqnarray}
[q_{\alpha,j}(t),\dot{q}_{\beta,k}^{\dag}(t)]=[q_{\beta,j}^{\dag}(t),\dot{q}_{\alpha,k}(t)]=i\hbar c^2\delta_{\alpha,\beta}\delta_{j,k}\label{51}
\end{eqnarray}
and $[q_{\alpha,j}(t),q_{\beta,k}(t)]=[q_{\alpha,j}(t),q_{\beta,k}^{\dag}(t)]=0$, $[\dot{q}_{\alpha,j}(t),\dot{q}_{\beta,k}(t)]=[\dot{q}_{\alpha,j}(t),\dot{q}_{\beta,k}^{\dag}(t)]=0$.
From these we deduce 
\begin{eqnarray}
[\mathbf{F}(\mathbf{x},t),\mathbf{\mathbf{\Pi}}_\mathbf{F}(\mathbf{x'},t)]=[\mathbf{F}(\mathbf{x},t),\mathbf{\mathbf{\Pi}}_\mathbf{F}^\dagger(\mathbf{x'},t)]\nonumber\\=\sum_{\alpha,\beta,j,k}\frac{[q_{\alpha,j}(t),\dot{q}_{\beta,k}^{\dag}(t)]}{c^2}\boldsymbol{\hat{\epsilon}}_{\alpha,j}\otimes\boldsymbol{\hat{\epsilon}}_{\beta,k}\Phi_\alpha(\mathbf{x})\Phi_\beta^\ast(\mathbf{x'})
\nonumber\\=i\hbar \boldsymbol{\delta_{\bot}}(\mathbf{x}-\mathbf{x'}).\label{52}
\end{eqnarray} with the unit transverse dyadic distribution:
\begin{eqnarray}
\boldsymbol{\delta_{\bot}}(\mathbf{x}-\mathbf{x'})=\sum_{\alpha,j}\boldsymbol{\hat{\epsilon}}_{\alpha,j}\otimes\boldsymbol{\hat{\epsilon}}_{\alpha,j}\Phi_\alpha^\ast(\mathbf{x'})\Phi_\alpha(\mathbf{x}).\label{53}
\end{eqnarray}
We have also $[\mathbf{F}(\mathbf{x},t),\mathbf{F}(\mathbf{x'},t)]=[\mathbf{\mathbf{\Pi}}_\mathbf{F}(\mathbf{x},t),\mathbf{\mathbf{\Pi}}_\mathbf{F}(\mathbf{x'},t)]=0.$\\
From these relations we deduce the commutation rules:
\begin{eqnarray}
[\mathbf{B}(\mathbf{x},t),\mathbf{B}(\mathbf{x'},t)]=[\mathbf{D}(\mathbf{x},t),\mathbf{D}(\mathbf{x'},t)]=0\label{54}
\end{eqnarray}
and \begin{eqnarray}
[\mathbf{B}_j(\mathbf{x},t),\mathbf{D}_k(\mathbf{x'},t)]=ic\hbar\sum_l\varepsilon_{j,k,l}\partial_l\delta^3(\mathbf{x}-\mathbf{x'})\label{55}
\end{eqnarray}
However, $X$ and $F$ conjugate variables commute as well and we have $[\mathbf{B}(\mathbf{x},t),\mathbf{P}(\mathbf{x'},t)]=[\mathbf{D}(\mathbf{x},t),\mathbf{P}(\mathbf{x'},t)]=[\mathbf{P}(\mathbf{x},t),\mathbf{P}(\mathbf{x'},t)]=0$. Therefore, from $\mathbf{E}(\mathbf{x},t)=\mathbf{D}(\mathbf{x},t)-\mathbf{P}(\mathbf{x},t)$ we deduce: $[\mathbf{E}(\mathbf{x},t),\mathbf{E}(\mathbf{x'},t)]=0$ and
\begin{eqnarray}
[\mathbf{B}_j(\mathbf{x},t),\mathbf{E}_k(\mathbf{x'},t)]=[\mathbf{B}_j(\mathbf{x},t),\mathbf{D}_k(\mathbf{x'},t)]\nonumber\\=ic\hbar\sum_l\varepsilon_{j,k,l}\partial_l\delta^3(\mathbf{x}-\mathbf{x'})\nonumber\\ \label{56}
\end{eqnarray} 
\indent Within this approach the  electromagnetic fields can also be directly expressed as a function of the rising and lowering operators and we get Eq.~\ref{57b}.
\indent Finally, In the configuration space  and by analogy with Eq.~\ref{59}, it is also possible to define the auxiliary `photon' fields  
\begin{eqnarray}
\mathbf{Z}^{(\pm)}(\mathbf{x},t)=\partial_t\mathbf{F}(\mathbf{x},t)\pm ic\sqrt{-\boldsymbol{\nabla}^2}\mathbf{F}(\mathbf{x},t)\nonumber\\ \label{61}
\end{eqnarray} which obey the following first order equations:
\begin{eqnarray}
\partial_t\mathbf{Z}^{(\pm)}(\mathbf{x},t)=\pm ic\sqrt{-\boldsymbol{\nabla}^2}\mathbf{Z}^{(\pm)}(\mathbf{x},t)+c^2\boldsymbol{\nabla}\times\mathbf{P}(\mathbf{x},t)\nonumber\\ \label{62}
\end{eqnarray}
The analogy with Eq.~48 allows us to introduce a photon field $\boldsymbol{\Psi}(\mathbf{x},t)$ such as
\begin{eqnarray}
\frac{\mathbf{Z}^{(-)}(\mathbf{x},t)}{2c}=\sqrt{\frac{\hbar c}{2}}(-\boldsymbol{\nabla}^2)^{1/4}\boldsymbol{\Psi}(\mathbf{x},t) \label{63}
\end{eqnarray}
with
\begin{eqnarray}
\boldsymbol{\Psi}(\mathbf{x},t)=\sum_{\alpha,j} c_{\alpha,j}(t)\boldsymbol{\hat{\epsilon}}_{\alpha,j}\Phi_\alpha(\mathbf{x}) \label{64}
\end{eqnarray}
The commutators  read now
\begin{eqnarray}
[\boldsymbol{\Psi}(\mathbf{x},t),\boldsymbol{\Psi}^\dag(\mathbf{x'},t)]= \boldsymbol{\delta_{\bot}}(\mathbf{x}-\mathbf{x'}) \label{65}
\end{eqnarray} with also $[\boldsymbol{\Psi}(\mathbf{x},t),\boldsymbol{\Psi}(\mathbf{x'},t)]=[\boldsymbol{\Psi}^\dag(\mathbf{x},t),\boldsymbol{\Psi}^\dag(\mathbf{x'},t)]=0.$
\section{Transformation matrix between $\textbf{A}$ and $\textbf{F}$ representations}
We have using Eq.~\ref{42}:
\begin{eqnarray}
-\sum_{\alpha,j} \frac{\dot{x}_{\alpha,j}(t)}{c}\boldsymbol{\hat{\epsilon}}_{\alpha,j}\Phi_\alpha(\mathbf{x})\nonumber\\+\int_{0}^{+\infty}d\omega\sqrt{\frac{2\sigma_\omega(\mathbf{x})}{\pi}}\mathbf{X}_{\bot,\omega}(\mathbf{x},t)\nonumber\\=\sum_{\alpha,j}i\frac{\omega_\alpha}{c} q_{\alpha,j}(t)\hat{\mathbf{k}}_\alpha\times\boldsymbol{\hat{\epsilon}}_{\alpha,j}\Phi_\alpha(\mathbf{x})\label{72}
\end{eqnarray}
Similarly for the magnetic field
\begin{eqnarray}
\sum_{\alpha,j}i\frac{\omega_\alpha}{c} x_{\alpha,j}(t)\hat{\mathbf{k}}_\alpha\times\boldsymbol{\hat{\epsilon}}_{\alpha,j}\Phi_\alpha(\mathbf{x})\nonumber\\=\sum_{\alpha,j} \frac{\dot{q}_{\alpha,j}(t)}{c}\boldsymbol{\hat{\epsilon}}_{\alpha,j}\Phi_\alpha(\mathbf{x})\label{73}
\end{eqnarray}
From Eqs.~\ref{72} and \ref{73} we thus deduce the equivalence relations: 
\begin{eqnarray}
i\frac{\omega_\alpha}{c} q_{\alpha,1}(t)=-\frac{\dot{x}_{\alpha,2}(t)}{c}+P_{\alpha,2}(t)\nonumber\\
-i\frac{\omega_\alpha}{c} q_{\alpha,2}(t)=-\frac{\dot{x}_{\alpha,1}(t)}{c}+P_{\alpha,1}(t)\nonumber\\
\dot{q}_{\alpha,1}(t)= -i\omega_\alpha x_{\alpha,2}(t)\nonumber\\
\dot{q}_{\alpha,2}(t)= i\omega_\alpha x_{\alpha,1}(t)\label{74}
\end{eqnarray} with $P_{\alpha,j}(t)$ given by Eq.~\ref{75}.
Equivalently, we can write Eq.~\ref{74} in terms of rising and lowering operators and deduce Eq.~\ref{76}. We can also rewrite Eq.~\ref{76}  using  the unitary transformation (Power-Zienau) \cite{Zienau,Cohen2,Craig,Power1983,Wooley,Raab} $T(t)=e^{i\int d^3\mathbf{x}\textbf{A}\cdot \textbf{P}/(c\hbar)}$ (with  $T^\dagger=T^{-1}$) such as:
\begin{eqnarray}
 a_{\alpha,j}(t)+\frac{1}{\sqrt{2\hbar\omega_\alpha}}P_{\alpha,j}(t)=Ta_{\alpha,j}(t)T^\dagger.\label{76b}
\end{eqnarray}
In, particular we have directly in the Power Zienau transformation \begin{eqnarray}T^{-1}(t)\textbf{D}(\textbf{x},t)T(t)=\textbf{E}_\bot(\textbf{x},t)\nonumber\\
T^{-1}(t)\textbf{B}(\textbf{x},t)T(t)=\textbf{B}(\textbf{x},t)\nonumber\\T^{-1}(t)[\textbf{p}_n(t)-e_n\frac{\textbf{A}(\textbf{x}_n,t)}{c}]T(t)=\textbf{p}_n(t)
\end{eqnarray} This transformation can be used in different ways for defining equivalent representations of the electromagnetic field as discussed in Refs.~\cite{Cohen2,Ackerhalt1984}.


\end{document}